\documentclass[review,12pt,authoryear]{elsarticle}
\usepackage{natbib}
\usepackage[utf8]{inputenc}
\usepackage{amsmath, amstext, amsthm, amsfonts, amssymb, mathrsfs}
\usepackage{lineno}
\usepackage{geometry}
\usepackage[pdfborder={0 0 0}]{hyperref}
\graphicspath{{figures/}} 
\usepackage{subfigure}

\usepackage{multicol}
\usepackage{float}
\usepackage{soul}
\usepackage[titletoc,title]{appendix}
\usepackage[section]{placeins} 
\usepackage[pdftex]{color}

\usepackage{booktabs}
\usepackage{nicefrac}

\usepackage[colorinlistoftodos]{todonotes}

\newcommand{\mr}{$(M,R)$}
\newcommand{\mrsys}{$(M,R)$ systems}
\newcommand{\ca}{Ca$^{2+}$}

\usepackage[cal=boondoxo]{mathalfa}

\usepackage{tikz}
\usetikzlibrary{arrows,fit, shapes,positioning}

\begin{document}

\begin{frontmatter}

    \title{An applied mathematician's perspective on Rosennean Complexity}

    \author[goett]{Ivo Siekmann}
    \ead{ivo.siekmann@mathematik.uni-goettingen.de}
    \cortext[cor1]{Corresponding author}

    \address[goett]{Institute for Mathematical Stochastics, Georg-August-Universit\"at G\"ottingen,\\Goldschmidtstra\ss{}e 7, 37077 G\"ottingen, Germany}

    \begin{abstract}
  The theoretical biologist Robert Rosen developed a highly original approach for investigating the question ``What is life?'', the most fundamental problem of biology. Considering that Rosen made extensive use of mathematics it might seem surprising that his ideas have only rarely been implemented in mathematical models. On the one hand, Rosen propagates relational models that neglect underlying structural details of the components and focus on relationships between the elements of a biological system, according to the motto ``throw away the physics, keep the organisation''. Rosen's strong rejection of mechanistic models that he implicitly associates with a strong form of reductionism might have deterred mathematical modellers from adopting his ideas for their own work. On the other hand Rosen's presentation of his modelling framework, \mrsys, is highly abstract which makes it hard to appreciate how this approach could be applied to concrete biological problems. In this article, both the mathematics as well as those aspects of Rosen's work are analysed that relate to his philosophical ideas. It is shown that Rosen's relational models are a particular type of mechanistic model with specific underlying assumptions rather than a fundamentally different approach that excludes mechanistic models. The strengths and weaknesses of relational models are investigated by comparison with current network biology literature. Finally, it is argued that Rosen's definition of life, ``organisms are closed to efficient causation'', should be considered as a hypothesis to be tested and ideas how this postulate could be implemented in mathematical models are presented.
    \end{abstract}


\end{frontmatter}

\section{Introduction}
\label{sec:intro}

When for the first time I heard about Robert Rosen's life-long quest
for the secrets of life, his theory of~\mrsys\ and his approach to
complexity I didn't quite know what to make of all this. There was an
obviously highly original idea for investigating a question which is
so hard to answer that it is, in fact, rarely asked: What is life?
Also the methods that Rosen used for his work, borrowed from the
highly abstract theory of categories, do not quite fit in the
classical arsenal of the applied mathematician's toolbox.  Could
category theory, an area of mathematics so abstract that, in fact,
even some of its pioneers referred to it as ``abstract nonsense'' be
successfully applied to a fundamental real-world question ``What is
life?'' which at the same time happens to be one of the hardest
scientific questions that one may possibly ask?\footnote{{Rosen's work
    on \mrsys\ is by no means the only application of category theory
    to the sciences. Best-known are perhaps applications in computer
    science--- two examples for textbooks are \citet{Pie:91a} and
    \citet{Bar:12a}---as well as mathematical physics
    \citep{Coe:11a}. A recent introduction to category theory with a
    view towards applications in the sciences by \citet{Spi:14a}
    underlines the the fact that the trend of category-theoretic ideas
    in science is increasing. But Rosen's work is one of the earliest,
    if not the earliest application of category theory outside
    mathematics.}}  That sounded interesting, very interesting,
indeed!

So I asked two questions that I usually ask myself when I hear about
something new and exciting to me in science:
\begin{enumerate}
\item Which of Rosen's ideas can I steal for my own work? (more about
  stealing later, see Section~\ref{sec:steal}!)
\item Do I believe Rosen's answers to his research questions ``What is
  life?'' and ``What is a complex system?''
\end{enumerate}

I will present my answers to these questions as my personal
perspective on Robert Rosen's work. The purpose of this is two-fold:
First, in my opinion, Rosen's highly original work deserves more
attention from the mainstream of mathematical biologists. Second, I
believe that Rosen's frustration that his ideas were not more widely
and openly accepted \citep{Mik:01a} is not completely
coincidental---there are important differences between Rosen's
theoretical concept of a model and the understanding of modelling
within the applied mathematics community. {These differences are on
  the one hand philosophical---Rosen demands that a model accurately
  represents the causal relationships between the elements of the
  system to be modelled (see Section~\ref{sec:abstraction}) whereas
  models typically built by applied mathematicians can be regarded as
  formal representations of a \emph{hypothesis} regarding a possible
  mechanism underlying the system behaviour (see
  Section~\ref{sec:appliedmaths}). On the other hand, Rosen applies
  mathematical notions, in particular, category theory, in a different
  spirit than most applied mathematicians would. This issue---which is
  related to Rosen's presentation of his ideas rather than the ideas
  themselves---is more important than it may look at a first glance
  because this difference in using mathematical tools may deter an
  audience with a mathematical background from Rosen's ideas
  (Section~\ref{sec:criticalCT}).} My presentation is based on the
original publications \citet{Rosen:58a, Rosen:58b, Rosen:59a,
  Rosen:71a, Rosen:73a, Rosen:91a} and \citet{Rosen:00a} but I will
most often refer to \citet{Rosen:72a} because this, in my opinion, is
the best summary of Rosen's early publications and to his monograph
``Life Itself'' \citep{Rosen:91a} which is the most comprehensive
account of the philosophical basis of Rosen's work. Another good
introduction into Rosen's thinking are his ``Autobiographical
Reminiscences'' \citep{Rosen:06a}.

The article is structured as follows: In Section~\ref{sec:life} we
introduce the notions of metabolism-repair systems (\mrsys). In
Section~\ref{sec:complex} we present Rosen's proposed characterisation
of life as systems that are ``closed to efficient causation''. We show
that this concept is not---as Rosen suggests---a specific property
that can be deduced from the architecture of~\mrsys\ but should be
regarded as a postulate, a hypothesis to be tested by implementing
``closure to efficient causation'' in mathematical models. Rosen's
specific view of modelling which is closely related to his
interpretation of category theory is presented in
Section~\ref{sec:abstraction}. I describe the conceptual basis of
mechanistic models in Section~\ref{sec:appliedmaths}. In particular, I
will argue that Rosen's relational models can be regarded as a
specific type of mechanistic models. In the Discussion
(Section~\ref{sec:discussion}) I compare mechanistic models with
Rosen's perspective on modelling and present some ideas how his
concept of an organism could be investigated via mathematical models
in physiology and ecology.

\section{Rosen's answer to the question ``What is life?''}
\label{sec:life}

Although most people---with or without a scientific background---seem
to have a good intuition when it comes to decide if something is
``alive'' it is nevertheless very hard to come up with a rigorous
scientific definition of life. 
Thus, definitions of life are usually descriptive---a list of
properties that are characteristic of living systems is given such as
the following appearing in \citet{Cam:08a}:
\begin{enumerate}
\item \emph{Order.} Organisms are highly ordered, and other characteristics of life emerge
from this complex organization.
\item \emph{Reproduction.} Organisms reproduce; life comes only from life \emph{(biogenesis)}.
\item \emph{Growth and Development.} Heritable programs stored in DNA direct the
species-specific pattern of growth and development.
\item \emph{Energy Utilization.} Organisms take in and transform energy to do work,
including the maintenance of their ordered state.
\item \emph{Response to Environment.} Organisms respond to stimuli from their
environment.
\item \emph{Homeostasis.} Organisms regulate their internal environment to maintain a
steady-state, even in the face of a fluctuating external environment.
\item \emph{Evolutionary Adaptation.} Life evolves in response to interactions between
organisms and their environment.
\end{enumerate}
But these properties are not necessarily defining: systems that are
not usually considered to be living systems may have one or even
several of these properties. Indeed, \citet{Cam:08a} refers to this
list as emergent properties and processes of life rather than a
definition.

Instead of a descriptive definition, Rosen proposes a relational
approach for distinguishing systems that are ``dead'' from systems
that are ``alive''. He starts from a set of components that he
explicitly refers to as black boxes i.e. he avoids making any
assumptions on the internal structure of these components. Instead his
focus is on the relationships between these components---he develops a
highly abstract theory with the purpose of demonstrating that the way
that components interact determines if a system is ``complex'' or
``simple'' and also, if a system is ``alive'' or ``not alive''. By
developing an approach that intentionally ignores the properties of
individual components of a system and emphasising the relationships
between these components he followed a motto of his mentor Nicolai
Rashevsky (cited according to \citet{Rosen:06a})---``Throw away the
physics, keep the organisation''.

More generally, the question of the relationship between structure
(i.e., for example, the underlying physics) and function in biology
has a long history. For example, the famous Cuvier-Geoffroy debate in
front of the French Academy of Sciences in 1830 was ultimately about
the two principles ``form follows function'' which was Georges
Cuvier's view whereas Geoffroy Saint-Hilaire argued for the opposite
position ``structure determines function''. In \citet{Rosen:91a}, his
monograph ``Life itself'', he strongly rejects ``structure determines
function'' which is currently, for example, influential in molecular
biology in the theory of protein folding---because the sequence of
amino acids (primary structure) to a great extent controls the
three-dimensional arrangement (tertiary structure) and this 3D
structure determines the function of a protein it is argued that
structure determines function \citep{Pet:08a}. 

In contrast, Rosen states that biological functions arise from the
interactions between the parts of a biological system, independent of
the material realisation of the components. In order to explain this
idea, let us consider calcium signalling. In many cases when hormonal
or electrical signals reach a cell, calcium oscillations are used for
propagating these signals within the cell and control a wide range of
cellular functions such as the contraction of heart cells or the
transcription of particular genes. The shape of these oscillations can
be very different between cell types although the \ca\ signalling
components involved are the same---voltage-gated \ca\ channels, that
allow calcium influx in response to electrical signals, intracellular
channels like the inositol-trisphosphate or the ryanodyne receptor,
that release large amounts of calcium from intracellular stores when
stimulated, and \ca\ pumps, that return \ca\ released to the cytosol
back to intracellular stores. How can \ca\ oscillations be so
different in different cell types if they are generated by similar
sets of \ca\ signalling components? An obvious explanation is that
\ca\ oscillations in particular cell types are shaped by relationships
between the components that are characteristic of this cell type. This
is the concept of the \ca\ ``toolbox'' which is the basis of our
current understanding of \ca\ oscillations \citep{Ber:00a}. But does
the fact that differences in the relationships between components are
important for explaining the different shape of \ca\ oscillations
imply that we should restrict ourselves to investigating relationships
and completely ignore structural properties of the components? We will
come back to this question in more general terms in the Discussion.

From Rosen's introduction of his~\mrsys\ it is quite clear that he not
only wishes to apply this approach for explaining the behaviour of
particular biological systems but from the outset he aims for
answering the grand question ``What is life?''. As for many models,
also for~\mrsys\ the answer to his question ``What is life?'' is
already determined to some extent by the construction of the
model---this will be explained in more detail in
Section~\ref{sec:appliedmaths}. 
Here, we will demonstrate that Rosen focuses mostly on two of the
aspects of life mentioned above, energy utilisation (in the following
referred to as metabolism) and homeostasis.

The~$M$ in \mrsys\ refers to metabolism. Metabolism is formally
modelled as the transformation of ``input materials'' to ``output
materials'' via the action of \emph{components}. Mathematically,
components are represented as mappings
\begin{equation}
  \label{eq:map}
  f: A \to B, \quad a \mapsto b=f(a)
\end{equation}
between sets~$A$ (``input materials'') and~$B$ (``output
materials''). In biochemical terms, $f$ may be interpreted as an
enzyme because it catalyses the transformation of elements of~$A$ to
elements of~$B$ while remaining unchanged.\footnote{An adaptation of
  Rosen's terminology to biochemistry may be found in
  \citet{Let:06a}. Here, we keep the original terminology.} However,
in real metabolic networks, enzymes degrade so that~$f$ will
eventually ``disappear''.\footnote{The nature of this
  ``disappearance'' is not made explicit in Rosen's writings. He only
  states that disappearance of a component has the effect that the
  production of output material~$B$ stops.}  Thus, in order to ensure
long-term stability, Rosen assumes for each component~$f$ of an
\mrsys\ the existence of a component~$\Phi$ that ``replicates''~$f$
should it be degraded. We denote such a component~$\Phi_f$ if we want
to emphasise the fact that~$\Phi_f$ replicates the component~$f$. The
mappings~$\Phi$ are again components (called \emph{repair components})
as defined in~\eqref{eq:map}
\begin{equation}
  \label{eq:repair}
  \Phi_f: C \to H(A,B), \quad c \mapsto \Phi_f(c)
\end{equation}
but the range of~$\Phi_f$ must be~$H(A,B)$, the set of maps
between~$A$ and~$B$.\footnote{{By using this notation, Rosen would
    like to imply that the maps appearing in~$H(A,B)$ are
    ``(homo)morphisms'', maps that preserve mathematical structure
    associated with~$A$ and~$B$ rather than general maps---see
    Section~\ref{sec:abstraction} for an explanation of the
    category-theoretic notion of morphism via an example. But because
    Rosen avoids assigning a specific mathematical structure to the
    sets~$A$ and~$B$ this has no consequences for his model, in
    particular, the maps~$H(A,B)$ cannot model, for example,
    biochemical properties of metabolism. Moreover, in many
    circumstances the morphisms~$H(A,B)$ are still just a set even
    if~$A$ and~$B$ have a particular mathematical structure.}}  Also,
it is postulated that the domain~$C$ of~$\Phi_f$ contains at least one
\emph{environmental output}~$O$ i.e. there exits a subset
of~$O\subset C$ that does not contain the domain of any component~$f$
\citep{Rosen:72a}.\footnote{By introducing~$\Phi_f$, strictly
  speaking, the set $O$ fails to be an environmental output because it
  is now contained in the domain of~$\Phi_f$.} With the repair
components~$\Phi$ (the~$R$ in \mrsys), Rosen adds a representation of
homeostasis to his \mrsys ---each $\Phi_f$ ensures the continuous
operation of a particular component~$f$. To give the repair components
a biological interpretation similar to the enzymes~$f$, Rosen
sometimes refers to the~$\Phi$ as genetic components. Thus, although
this is not stated explicitly in Rosen's writings, he therefore
characterises life as the combined effect of metabolism and
homeostasis.\footnote{In Rosen's own biochemical interpretation of the
  components~$f$ as enzymes and~$\Phi$ as genetic components, \mrsys\
  can be related to the production of enzymes via gene transcription
  which itself depends on the activity of enzymes. But the abstract
  construction of \mrsys\ are general enough that the approach can be
  applied to different domains.}

Unfortunately, whereas the degradation of components~$f$ can be
prevented by the repair components~$\Phi$, repairing these newly
introduced components~$\Phi$ would require another set of repair
components. But Rosen was able to demonstrate that the infinite
regress of having to add more and more repair components could be
avoided---he proposed that under certain assumptions a repair
component for~$\Phi_f$ could be identified with an element from the
range of~$f$. Because a repair component for~$\Phi_f$ (which Rosen
denotes~$\beta$ and names \emph{replication map}) therefore does not
need to be added to the system, the infinite regress is avoided. We
will discuss the construction of~$\beta$ in the next section, Rosen
himself explains the details most clearly from a a mathematical point
of view in \citet{Rosen:72a}.

In summary, we observe that by suitably combining metabolic and repair
components, an~\mrsys\ is capable of achieving homeostasis. Although
all components have a limited life time, the system is able to survive
for much longer (in theory indefinitely) because components are
replaced early enough before they degrade. Of course, this only refers
to the components that are parts of the~\mrsys\ because at least some
of them depend on environmental inputs. This demonstrates that~\mrsys\
are able to autonomously maintain their internal organisation,
provided that an ``energy source'' (via environmental inputs) is
available. We will explain Rosen's own formulation of this result in
more detail in the next section.

\section{Closure to efficient causation}
\label{sec:complex}

In the previous section we explained that through interactions of
metabolic and repair components a \mr~system achieves some level of
autonomy---it is capable of maintaining its internal components (which
all have a limited life time) by drawing on an energy source from the
environment. Rosen summarised this this as ``organisms are closed to
efficient causation''. In ``Life Itself'' he discusses in detail how
his theoretical ideas relate to the four Aristotelean
causes,\footnote{Applications of Aristotle's classification of four
  causes have some tradition in biology, confer Tinbergen's levels of
  analysis, first presented in \citet{Tin:63a}.} one of which is the
\emph{efficient cause}. The efficient cause is most closely related to
the modern notion of causality. In the context of~\mrsys\ each of the
components are the efficient causes of the transformation of elements
in the domain to elements in the range, for example,
in~\eqref{eq:map}, $f$ is the efficient cause for transforming~$A$
to~$f(A) \subset B$. For the same example, $A$ provides the
\emph{material cause} for the transformation of~$A$
to~$f(A)$.\footnote{Rosen also discusses the \emph{formal cause} and
  the \emph{final cause}. Although all four causes are important for
  Rosen's theory, the remaining two Aristotelean causes are not
  directly relevant to our discussion. Therefore we refer the reader
  to \citet{Rosen:91a}.}

We only consider the simplest~\mr\ system,\footnote{This is, in fact,
  not as strong a restriction as it may seem---by combining sets via
  Cartesian products~`$\times$' and defining maps on the products in
  an obvious way, several~\mrsys\ with more than three components can
  be cast in the form~\eqref{eq:ABC}.} consisting of one metabolic
component~$f$ and one repair component~$\Phi_f$. In order to avoid
degradation of the repair component~$\Phi_f$ we need an additional
repair component~$\beta$ that
replaces~$\Phi_f$. Summarising~\eqref{eq:map}, \eqref{eq:repair} and
adding~$\beta$ we get
\begin{equation}
  \label{eq:ABC}
  A \xrightarrow{f} B \xrightarrow{\Phi_f} H(A,B) \xrightarrow{\beta} H(B, H(A,B)).
\end{equation}
Rosen develops a complicated argument why the map~$\beta$
in~\eqref{eq:ABC} can be identified with an element~$b$ of the
set~$B$. He explicitly constructs a parameterisation 
\begin{equation}
  \label{eq:parameterB}
  B \ni b \mapsto \beta_b: H(A,B) \to H(B, H(A,B))
\end{equation}
that assigns a map~$\beta_b$ to each~$b \in B$. Rosen's interpretation
of~\eqref{eq:parameterB} is that each~$\beta_b$ is, in fact, already
contained in set~$B$ and needs not be explicitly added to the
system. Rosen's construction of the~$\beta_b$ has caused a
considerable amount of confusion---some authors disputed its
mathematical feasibility \citep{Lan:02a} whereas others responded to
this claim by explicitly constructing sets and maps as
in~\eqref{eq:ABC} following Rosen's approach \citep{Let:06a}.

In contrast, I will argue that from a mathematical point of view there
is, in fact, nothing to show. The only restriction for~$\beta$ is the
condition that~$\beta(f)=\Phi_f$ which means in terms of~\mrsys\
that~$\beta$ repairs~$\Phi_f$ by transforming~$f$. After choosing any
map~$\beta$ that fulfils this condition, a parameterisation~$\beta_b$
as in~\eqref{eq:parameterB} can be easily obtained---we only need to
select an arbitrary~$b^*$ that maps to~$\beta_{b^*}=\beta$ and map all
other elements of~$B$ to arbitrary~$\beta_b$.

Rosen's difficulty arises because he insists on deriving the
parameterisation~$\beta_b$ from evaluation maps, see \citet{Rosen:72a}
where~$\epsilon_b$ is denoted~$\hat{b}$:
\begin{equation}
  \label{eq:evaluateB}
  \epsilon_b: H(B, H(A,B)) \to H(A,B), \quad \Phi \mapsto \Phi(b).
\end{equation}
But there is no reason for constructing the maps~$\beta_b$ in this
way---the fact that the~$\beta_b$ were obtained from evaluation
maps~$\epsilon_b$~\eqref{eq:evaluateB} never plays a role in Rosen's
discussions of the replication map~$\beta$. In summary, as Rosen
postulated, it is indeed possible to construct~\mrsys\ where all
components are ``repaired'' by other components once they have
surpassed their finite life time. But the realisation of the
map~$\beta$ does not, as Rosen suggests, follow from the architecture
of~\mrsys. Instead, I propose to consider the existence of the
replication map simply as a postulate regarding the structure of
living systems, summarised in the statement ``organisms are closed to
efficient causation''.

\begin{figure}[htbp]
  \centering
  \begin{tikzpicture}
    \node(A){A};
    \node[right=of A, shape=circle](B){B}; 
    \node[right=of B](phi){$\Phi_f 
      $};
    \node[above=of B,shape=circle](f){$f 
      $};
    \node[above=0.3cm of f](Hab){$H(A,B)$};

    \path (f) -- node[sloped] {$\in$} (Hab);




    \node[below= 0.25cm of B](b){$b$};
    \path (b) -- node[sloped] {$\in$} (B);

    \node[right= 0.5cm of b](beta){$\beta_{b} 
      $};
    \path (beta) -- node{$\mapsto$} (b);

    \node[right=0.3cm of phi](HbHab){$H(B, H(A,B))$};
    \path (phi) -- node{$\in$} (HbHab);

    \path (f.south west) +(4pt,0) node (fbl){} (f.south east) +(-4pt,0) node (fbr) {};
    \path (B.north west) +(4pt,0) node (Bul){} (B.north east) +(-4pt,0) node (Bur) {};

    \draw[->,>=triangle 45](A) -- (B);
    \draw[->,>=triangle 45] (Bul) -- (fbl);
    \draw[->,>=triangle 45] (f) -- (phi);

    \draw[->,>=
    triangle 45,dotted](phi) -- (B);
    \draw[->,>=
    triangle 45, dotted] (Bur)  -- (fbr);
    \draw[->,>=
    triangle 45, dotted] (f) -- (A);
  \end{tikzpicture}
  \caption{\emph{Closure to efficient causation:} This diagram
    (adopted from \citet[chapter 10]{Rosen:91a}) illustrates the
    different processes that the individual components of the~\mr\
    system in~\eqref{eq:ABC} are involved in. On the one hand, solid
    arrows show where transformations from input material to output
    material occur---~$A$ is transformed to~$B$, $B$ is, in turn used
    for ``repairing''~$f$ whereas $f$ is transformed in order to
    repair~$\Phi_f$. On the other hand, for the broken arrows, a
    component located at the start of an arrow indicates the
    initiation of a transformation located at the arrow tip---$f$
    catalyses the ``metabolic'' transition of~$A$ to~$B$, $\Phi_f$
    acts on~$B$ for repairing~$f$ and the replication map~$\beta_b$
    starts the repair of~$\Phi_f$ by transforming~$f$. It is clear
    that each of the components~$f$, $\Phi_f$ and~$\beta_b$ regulates
    the repair of another component. In Rosen's words, the sets~$A$,
    $B$, $H(A,B)$ and~$H(B, H(A,B))$ can be regarded as ``material
    causes'' whereas $f$, $\Phi_f$ and~$\beta_b$ are ``efficient
    causes''. Because each of the components is in turn produced by
    one of the other components the system is \emph{closed to
      efficient causation}.}
  \label{fig:closure}
\end{figure}
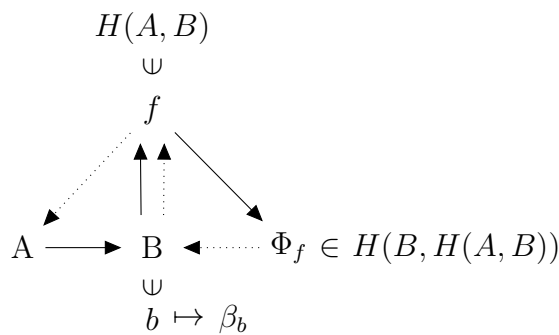

More interesting than the details of the mathematical investigation is
Rosen's interpretation of~\mrsys\ whose elements interact in a way
that mutually ensures replacement of failing components. \mrsys\ with
this property provide Rosen's model for organisms which he
characterises as ``organisms are closed to efficient causation''. In
order to explain this concept, \citet[chapter 10]{Rosen:91a} presents
a diagram that illustrates this statement~(Fig.~\ref{fig:closure}).
The diagram shows---indicated by solid arrows---the ``material''
transformations between elements of the sets~$A$, $B$, $H(A,B)$,
$H(B, H(A,B))$ and~$H(H(A,B), H(B, H(A,B))$ (by a mapping to~$B$) but
also shows ``causal'' relationships---indicated by broken
arrows---where components initiate a transformation by acting on
elements in one of these sets. By following a broken and then a solid
arrow we see that the production of each component is ``caused'' by
another component in the system---in Rosen's own words, $f$
``entails''~$\beta_b$, $\beta_b$ ``entails''~$\Phi_f$ and~$\Phi_f$
``entails''~$f$. Thus, the system contains a closed loop of efficient
causation~$f\to\beta_b\to\Phi_f\to f$, a property that Rosen
denotes~\emph{closure to efficient causation}.

In summary, in this section we have explained Rosen's proposed
definition of living systems,``organisms are closed to efficient
causation''.  We have shown that ``closure to efficient causation'' is
not a result that follows from the construction of~\mrsys. In
contrast, the ability of organisms to autonomously maintain their
internal organisation should be regarded as a postulate, a hypothesis
to be tested for concrete biological systems by developing
mathematical models. We will return to this important idea in the
Discussion. 

\section{The modelling relation}
\label{sec:abstraction}

Interesting about Rosen's approach are not only \mrsys\ themselves but
also how they are constructed and investigated. He uses category
theory, one of the most abstract mathematical disciplines, that was
developed starting from the 1940s.\footnote{The following introduction
  to category theory is intentionally informal---the point is to
  introduce the spirit of category theory rather than enabling the
  reader to start their own career as a category theorist. See the
  Introduction of the classical monograph by \cite{Mac:71a} for a very
  readable exposition for the mathematically inclined reader.} In
order to give a simple example that introduces many important aspects
of category theory without requiring much mathematical background
consider a planar algebraic curve.\footnote{This is, in fact, the
  first example of category theory that I saw as a student. I thank
  Prof. Heinz Spindler (University of Osnabr\"uck, Germany) for his
  beautiful lectures on algebraic geometry that gave me a lot of
  pleasure.}  Readers that are familiar with category theory may
safely skip this slightly lengthy example but I hope that it may be
helpful for readers without prior knowledge of category theory. The
main purpose of this section is to explain that the precise meaning of
Rosen's notion of a \emph{modelling relation} may be regarded as the
\emph{representation} of a \emph{functor} between different
categories. This has been explained more formally by \cite{Lou:09a}.

\subsection{Algebraic curves or: an introduction to category theory}
\label{sec:algebraiccurves}

An algebraic curve in the plane is defined by a polynomial equation in
two indeterminates~$x$ and~$y$ i.e. $f$ is an element of a polynomial
ring~$K[x,y]$. Examples include the parabola, $y-x^2=0$ or the unit
circle, $x^2 + y^2 = 1$. Geometrically, the curve is the set of
pairs~$(a,b)$ that fulfil the equation~$f(x,y)=0$; the $(a,b)$ are
elements of a two-dimensional vector space~$R^2$. One of the most
important aspects of modern algebraic geometry is not to be very
specific about~$R$, the interest is in general properties of algebraic
sets defined by systems of polynomial equations. Whereas classical
algebraic geometry investigates polynomial equations over the complex
numbers~$\mathbb{C}$, we may instead consider the real
numbers~$\mathbb{R}$, the rationals~$\mathbb{Q}$ or a finite field
with~$q$ elements~$\mathbb{F}_q$.\footnote{A \emph{field}~$K$ is an
  algebraic structure where addition and multiplication of elements
  are \emph{associative}, \emph{commutative} and \emph{distributive}
  and additive and multiplicative inverses exist. This means that for
  each~$a \in K$ we find a~$b \in K$ so that~$a+b=0$ ($b$ is
  denoted~$-a$) and for each~$c\in K$, $c \neq 0$ there is a~$d\in K$
  such that~$c \cdot d=1$ ($d$ is denoted~$d^{-1}$). Examples are the
  complex numbers~$\mathbb{C}$, the real numbers~$\mathbb{R}$, the
  rational numbers~$\mathbb{Q}$ or finite fields~$\mathbb{F}_q$
  with~$q$ elements.} Even more generally, we may choose an
algebra~$R$ over a field~$K$. A~$K$-algebra is a vector space over a
field~$K$ whose elements can in addition be multiplied, unlike in a
general vector space which only requires addition of
elements. Because, trivially, a field can be regarded as a
one-dimensional~$K$-algebra over itself, this includes the examples of
different fields given above. The generality with which~$R$ can be
chosen is expressed quite clearly by the following category-theoretic
description of an algebraic curve. For an arbitrary non-constant
polynomial~$f \in K[x, y]$, the affine planar algebraic curve~$C$
over~$R$ with equation~$f$ is defined as the \emph{functor}
\begin{equation}
  \label{eq:curvefunc}
  \mathcal{C}: \mathcal{Alg}_K \to \mathcal{Sets}
\end{equation}
with
\begin{equation}
  \label{eq:curveobj}
  \mathcal{C}(R) = \{ (a,b) \in R^2 | f(a,b) = 0\}
\end{equation}
for all~$K$-algebras~$R$. Thus, $\mathcal{C}(R)$ is the set of zeroes
of~$f$ over~$R^2$ which means that the functor~$\mathcal{C}$ provides
us with a set for each~$K$-algebra. From the category-theoretical
point of view, each $K$-algebra~$R$ is an \emph{object} of the
category of ``all''~$K$-algebras~$\mathcal{Alg}_K$ and ``all'' sets
are the objects of the category~$\mathcal{Sets}$\footnote{Due to
  set-theoretic paradoxes, we can, in fact, not consider categories of
  ``all'' $K$-algebras or ``all'' sets.}. But a category not only
consists of objects but also of maps between objects that preserve the
algebraic structure of the objects, so-called
\emph{morphisms}. For~$\mathcal{Sets}$ the morphisms are just ordinary
maps but for~$\mathcal{Alg}_K$ these are~$K$-algebra
homomorphisms~$\phi: R \to S$. As a map between categories, a functor
not only relates objects but also morphisms. For our example, for
a~$K$-algebra homomorphism~$\phi: R \to S$ between~$K$-algebras~$R$
and~$S$ we obtain a corresponding map
~$\phi_{\ast}:\mathcal{C}(R) \to \mathcal{C}(S)$ between the
sets~$\mathcal{C}(R)$ and~$\mathcal{C}(S)$ via
\begin{equation}
  \label{eq:evaluationmap}
  \phi_{\ast}:\mathcal{C}(R) \to \mathcal{C}(S), \quad (a,b) \mapsto (\phi(a), \phi(b)).
\end{equation}
Because~$\phi$ is a~$K$-algebra homomorphism it is indeed true that
if~$(a,b)$ is a zero of~$f$ (over~$R^2$), $(\phi(a), \phi(b))$ is also a
zero of~$f$ (over~$S^2$).

This abstract representation of a planar algebraic curve as a
relationship between~$K$-algebras and the geometric objects of
interest, the sets~$\mathcal{C}(R)$, has a punch line that may give
some insight why category-theoretical ideas have been quite successful
in some areas of mathematics but, more importantly, clarify one of
Rosen's key ideas, the \emph{modelling relation}. It can be shown that
for describing algebraic curves it is sufficient to consider only one
particular~$K$-algebra, the \emph{coordinate ring}\footnote{A
  (commutative) ring is an algebraic structure similar to a field. The
  difference is that multiplicative inverses are not required to exist
  for all elements. Note that in addition to the ring structure the
  coordinate ring also has the structure of a~$K$-algebra.}
\begin{equation}
\label{eq:coordinatering}
A = K[x,y]/(f), \quad \text{where } (f)=f K[x,y].
\end{equation}
With the expression $f K[x,y]$ we denote the set of all polynomials
that contain the polynomial~$f$ as a factor. The crucial point is that
the the coordinate ring~$A$ alone is sufficient for finding the
curves~$\mathcal{C}(R)$ for \emph{all}~$K$-algebras~$R$. We remind the
reader that the functor~$\mathcal{C}$ assigns the evaluation
map~$\phi_*: \mathcal{C}(A) \to \mathcal{C}(R)$ defined
in~\eqref{eq:evaluationmap} to each~$K$-algebra
homomorphism~$\phi: A \to R$. It can be shown that we can calculate
the curve~$\mathcal{C}(R)$ via
\begin{equation}
  \label{eq:evalfunctor}
  \phi_{*}([x], [y]) = (\phi([x]), \phi([y])), \quad [x], [y] \in A.
\end{equation}
by evaluating~$\phi_{*}$ at the particular
point~$([x],[y])$.\footnote{$[x]$, $[y]$ are equivalence classes and
  can be considered as those polynomials with ``remainder''~$x$
  or~$y$, respectively, when ``dividing'' by~$f$---the coordinate ring
  is an example of a \emph{quotient ring}. We will not go into more
  detail because it does not add much to the discussion and refer the
  interested reader to any introduction to algebra.} In
category-theoretical terms, the coordinate ring~$A$ is called a
\emph{representation} of the functor~$\mathcal{C}$ via the
\emph{universal element}~$([x], [y])$. Informally this means in the
context of our example that some objects of the
category~$\mathcal{Sets}$, algebraic curves defined by~$f$, can be
understood by analysing a particular~$K$-algebra, the coordinate
ring~$A$.

\subsection{The modelling relation and simulation}
\label{sec:modelling}

We will now---still with the example of a planar algebraic
curve---explain that Rosen's idea of a \emph{model} can be understood
as a functor between a natural and a formal system. Both the natural
as well as the formal systems are represented as categories, in fact,
shortly after the initial introduction \citep{Rosen:58a},
\citet{Rosen:58b} redefined \mrsys\ using category-theoretical
ideas{---we will discuss Rosen's use of category theory in
  Section~\ref{sec:categorytheory}}. It is instructive to observe how
the representation of the ``planar algebraic curve
functor''~$\mathcal{C}$ that assigns to a~$K$-algebra~$R$ a
curve~$\mathcal{C}(R)$ is used by mathematicians working in the field
of algebraic geometry.

Let us say that algebraic curves are the ``natural system'', graphs
obtained by finding the zeroes of a polynomial equation~$f(x,y)=0$. In
contrast, we regard~$K$-algebras as a ``formal system''
that---according to Rosen---encodes the ``causal entailments'' present
in the natural system. In general, causal entailment refers to the
relations between objects defined by the morphisms of a category which
is quite abstract. But for the specific example of algebraic curves it
is quite clear what this means. From a mathematical point of view,
studying a general algebraic curve over a general~$K$-algebra~$R$ is
impossible without further assumptions on the~$K$-algebra~$R$ because
sets do not have a lot of structure. In contrast, for~$K$-algebras,
arithmetical operations such as addition and multiplication are
defined and mathematical theorems have been proven that give insight
into how the elements behave under these operations. Thus, whereas the
structure of~$K$-algebras may be investigated with a variety of tools
from commutative algebra, much less insight can be gained by simply
considering the sets~$\mathcal{C}(R)$ of algebraic curves
over~$R$. But the functor~$\mathcal{C}$ enables us to switch between
the ``natural system'' (sets) and the ``formal system'' ($K$-algebras)
so that we can explore geometric facts using algebra. Incidentally,
algebraic geometers are well aware of this and refer to this process
with the motto ``Think geometrically, prove algebraically!''
\citep{Ale:91a}.

Each~$K$-algebra~$R$ can be understood as a ``model'' of
another~$K$-algebra~$S$ because we can ``translate'' the algebraic
curve~$\mathcal{C}(R)$ over~$R$ to the algebraic
curve~$\mathcal{C}(S)$ over~$S$ using the functor~$\mathcal{C}$ by
assigning the evaluation map~$\phi_*$ to each~$K$-algebra
homomorphism~$\phi: R \to S$.  Even better, because the
functor~$\mathcal{C}$ has a representation via the coordinate
ring~$A$~\eqref{eq:coordinatering} we may even resort to studying only
\emph{one}~$K$-algebra, namely the coordinate ring, and translate the
results to any other~$K$-algebra via evaluating the functor at the
universal element~\eqref{eq:evalfunctor}.

Whereas the functor~$\mathcal{C}$ provides us with an example of a
\emph{model} in Rosennean terms, we may also look for an example
of~\emph{simulation} in the same context. Rosen defined a simulation
as a relationship that considers the natural system as a ``black box''
without attempting (or being able) to capture the ``causal
entailment'' within the natural system. An example of simulation for
our example is the numerical approximation of algebraic
curves. Numerical methods may succeed in obtaining an approximation of
an algebraic curve without any consideration of the underlying
algebraic structure by iteratively approximating points of the curve
from a starting value~$(x_0, y_0)$ known to lie on the curve
\citep{Gom:09a}. With Rosen we might say that these numerical methods
are able to ``predict'' the ``natural system'' i.e. the algebraic
curve. But it is clear that this is not based on bringing the
entailment structure of a formal system in congruence with the
entailment structures of the natural system to be modelled. This,
however, is according to Rosen the ideal that a model should live up
to. \footnote{{An anonymous reviewer brought to my attention that two
    famous articles by Alan Turing provide a good example for the
    difference between simulation and model. The imitation game (also
    known as the Turing test) was proposed by \cite{Tur:50a} in order
    to answer the question ``Can machines think?''. In order to pass
    the test the machine must communicate in natural language with a
    human evaluator and through this conversation convince the
    evaluator that it is human. This is a perfect example for
    simulation because by definition of the test it is unimportant if
    and to which extent the machine attempts to accurately represent
    human intelligence. In contrast, \citet{Tur:52} proposes a
    mathematical model that exhibits inhomogeneous stationary
    distributions (Turing instability). This leads to a model that
    explains morphogenesis in terms of two interacting chemical
    species (an ``activator'' and an ``inhibitor'') that diffuse with
    different speeds.}} As I will explain in the Discussion, it is
exactly this ambition that, in my opinion, separates Rosen's
understanding of modelling most strongly from an applied
mathematician's view on modelling which is illustrated in the next
section.

\section{Lie, cheat and steal---the applied mathematician's ways for
  finding the truth}
\label{sec:appliedmaths}

I believe that there is no elaborated philosophy of modelling in
applied mathematics that could be compared to Rosen's. So it might be
helpful to first look at \emph{Dynamic Models in Biology}, a highly
readable introduction to mathematical biology by \citet{Ell:06a}. Near
the end of the book the authors briefly introduce the ``three
commandments of modelling'':
\begin{enumerate}
\item Lie
\item Cheat
\item Steal
\end{enumerate}

After explaining the three commandments in a bit more detail I will
provide a description of models in applied mathematics that will be
compared with Rosen's modelling relationship in
Section~\ref{sec:discussion}.

\subsection{Lying}
\label{sec:lying}

Everyone knows that models are based on assumptions. What not everyone
knows is that models (at least the good ones \citep{Ell:06a}) are
based on \emph{false} assumptions. As illustrations we can take nearly
all mathematical models from physics. One of the most striking example
is the apparently harmless notion of a mass point. Moving bodies such
as cars, space ships or parachutists are described by particles
without spatial extension whose mass is concentrated in a single
point. 
In this way our model describes objects that may weigh several tonnes
or more while at the same time they ``are not even there'' because it
is assumed to have no spatial extension! The only reason that we find
such an idea not completely outrageous is by the justification we get
for this model after we have applied it to a natural system. We ``get
away'' with this obviously wrong assumption, we can, for example,
predict the trajectories of celestial bodies to a certain
accuracy. Also, more detailed models of rigid body motion and the
notion of the centre of mass give additional support for this model
and insight why representing bodies as mass points worked in the first
place. The most important point here is, though, that before this
model was applied to a concrete problem, it was not at all clear that
it would turn out to provide a useful description of a physical
object. The justification of ``lies'' in modelling can only be given
in hindsight.

\subsection{Cheat}
\label{sec:cheat}

With cheating, \citet{Ell:06a} mostly refer to a particular way of
using statistics. They recommend to do things that ``would make a
statistician nervous'' by stretching the limits within which
statistical methods can be used instead of just ``letting the data
speak for itself''. It is not easy to provide an example for
``cheating'' because obviously scientists will usually not describe
anything they did in a study as cheating. Because I would not like to
accuse colleagues of cheating either I have no choice but to give an
example of ``cheating'' from my own work. A few years ago I was
working on a model for an ion channel \citep{Sie:12a}. My motivation
was to take into account model gating, a feature that is quite common
in ion channel dynamics but which has rarely been accounted for in
models. Instead of continuously adjusting their activity many ion
channels switch spontaneously between highly different types of
behaviour (modes). I wanted to demonstrate that across all
experimental conditions each of the different modes defined the same
type of behaviour i.e. could be described by statistically similar
models. Unfortunately, for some experimental conditions I was not able
to fit a model to the segments representative for the modes because
the channel was switching too fast and therefore the segments were too
short in order to produce statistically conclusive results. But
although I thus was not able to rigorously prove my claim I
nevertheless argued that the hypothesis of modes which are unchanged
across all experimental conditions was---with some positive and in the
absence of contrary evidence---presumably correct.

This way of using statistical methods emphasises that experimental
data is only one of many sources of knowledge that are synthesised in
a mathematical model, thus, the fate of a model should not depend
solely on the success or failure of a particular statistical method.

\subsection{Steal}
\label{sec:steal}

Although it may sound even worse than lying and cheating, in
scientific terms, stealing might actually be the most acceptable of
the three commandments. It simply means reusing ideas of models that
have previously appeared in the literature by, for example, applying
them to new systems. A famous example is the well-known Lotka-Volterra
model which can be seen as the beginning of predator-prey
modelling. \citet{Vol:26a} simply reinterpreted the law of mass action
kinetics where the rate of a chemical reaction is assumed proportional
to the product of the concentrations of the two reactants as the catch
rate of a predator feeding on a prey. The reason that terms
characteristic of chemical models were often ``stolen'' by ecologists
and epidemiologists is because the law of mass action and enzyme
kinetics can, from a more abstract point of view, be interpreted as
contact rates between two populations \citep{Sie:09a}.

We briefly mention one danger of stealing---in comparison to the
original domain of application, a ``stolen'' model may increase the
amount of lying and cheating---on the one hand the assumptions of the
original model may be less valid and on the other hand experimental
validation of the original model may not be available in the new
context.


\subsection{What is a mechanistic model?}
\label{sec:model}

In order to compare the approach followed in typical models in applied
mathematics with Rosen's modelling relation I will give a brief
description of models in applied mathematics. I will refer to these as
``mechanistic'' in the following because Rosen presumably means
similar models when he refers to mechanistic models. However, I will
argue in the Discussion that at least some of his objections only
arise when mechanistic models are interpreted in a reductionistic
sense.

The aim of a mechanistic model is to provide insight into a natural
system by synthesising different sources of knowledge. This is
achieved by defining a formal system that transparently captures how
the elements of the system interact with each other and how these
interactions are parameterised with experimental data. A mechanistic
model is also strongly determined by a purpose which means that
already the architecture and not just the interpretation of the model
results is defined by the question that the model should answer.

Building a mechanistic model starts with the formulation of a set of
assumptions that summarise what is known about a system, extended by
some hypotheses regarding details of the natural system that are
currently unknown. Which aspects of our knowledge are represented in
how much detail depends on the model purpose. Based on the underlying
assumptions a model structure is constructed that is meant to
represent the assumptions as well as possible. This model can then be
simulated in order to produce results. The results obtained from the
model are then interpreted in comparison with the natural system and
in the light of the assumptions made in the beginning which leads to
conclusions that are drawn from the modelling study.

To give an example, I will refer to recent work in mathematical
ecology that I contributed to. A series of papers, starting with
\citet{Ben:16a} and \citet{Sie:16b}, primarily had two purposes. The
aim of \citet{Ben:16a} was to investigate alternatives to the
classical model of population dispersal based on the common model for
diffusion due to Fick \citep{Fic:55}. \citet{Sie:16b} considered
alternatives to the classical model for environmental fluctuations
based on stochastic terms that scale linearly in the population
densities. \citet{Ben:16b} studied the combined effect of
Fokker-Planck diffusion \citep{Fok:14, Pla:17}, an alternative to
Fickian diffusion, and linear noise terms whereas \citet{Sie:17a}
combine Fokker-Planck diffusion with a nonlinear noise term proposed
in \citet{Sie:16b}. Consistent with the model purpose, the authors
primarily consider a very simple model for the interactions of
populations, the Lotka-Volterra competition model. Also, a special
parameter set is considered where in the deterministic, non-spatial
version of the model, the population with the higher initial
population always outcompetes its competitor. In order to investigate
spatial and stochastic effects, the parameter of the diffusion model
and the noise model are varied. The models demonstrate that stochastic
fluctuations enhance the success of invading species that invade the
habitat of a resident population but may also enable resident and
invader to coexist which is impossible in the deterministic
non-spatial version of the model.

At this point it is important to note that the definition of a
mechanistic model presented here by no means excludes relational
models favoured by Rosen---in the terms developed here, an important
underlying assumption of a relational model simply is that the
internal structure of the individual system elements is not
represented in detail. The most important difference to Rosen's
interpretation of models is that no suggestion is made that a
mechanistic model accurately captures the causal relationships of the
system to be modelled---there is no modelling relationship that in a
theoretical sense provides the modeller with access to the structure
of the natural system. The reason for this is scepticism---a large
proportion of applied mathematicians would presumably be highly
pessimistic that achieving congruence between a formal system and a
natural system as envisaged by Rosen was possible and even if it could
be achieved that this could be verified. Thus, from the outset, the
motivation of a mechanistic model is much more modest. The aim of
modelling is to provide insight into an \emph{aspect} of a natural
system that is defined by the \emph{purpose} of the model. Obtaining a
complete understanding is, in principle, out of reach due to the
simplifying assumptions made when the model was built. Also,
mechanistic models can only provide \emph{possible explanations} of
phenomena observed in the natural system. If the results obtained from
the model contradict the behaviour of the natural system one concludes
that the underlying assumptions of the model are either incorrect or
incomplete. But if the results are consistent with the system to be
modelled we cannot conclude that the explanation provided by the model
is correct because we cannot exclude the possibility that alternative
models with completely different underlying assumptions produce
similar results. Instead of regarding a mechanistic model as a
mathematical representation of some ``truth'' it is therefore more
accurate to think of a model as an argument for a particular
\emph{hypothesis} explaining the observed behaviour of a natural
system.

Finally, it is widely accepted among applied mathematicians that not
the development of individual models but the comparison of several
competing models of the same system that are based on different
assumptions provides most insight. Modelling is therefore not so far
from studying a system via experiments---with the important advantage
described by the mathematician Vladimir Arnold with the words
``mathematics is the natural science where experiments are cheap''.
This is very well illustrated by several monographs on mathematical
biology for which we give a few examples---the general introductions
by \citet{Mur:02, Mur:03}, \citet{Ede:05} and \citet{Ell:06a}
mentioned above, also we refer to more specific books on ecology
\citep{Oku:01,Mal:08a} or physiology \citep{Kee:09a, Kee:09b}.

\section{Discussion}
\label{sec:discussion}

\subsection{Rosen's answer to his question ``What is Life?''}
\label{sec:discussWhat}

As explained above, Robert Rosen looked for an answer to the question
``What is life?'' in a way that was different to the commonly used
modelling approaches at his time. Following the motto ``throw away the
physics, keep the organisation'' he proposed to investigate
relationships between ``components'' without associating these
abstract entities themselves with any structure.

Looking at this idea from the point of view of an applied
mathematician, we observe that Rosen's model starts with the ``lie'',
in the sense of Section~\ref{sec:lying}, that the physics of
components that make up a biological organism is mostly irrelevant for
understanding its functioning. A problem with this is that we can only
learn in hindsight, once this model has been applied to a real
biological organism, if this assumption has been able to provide new
insights. But because Rosen himself was more interested in developing
his formal framework and developing the theoretical ideas he drew from
those studies, his publications contain at most hints to possible
applications.

The idea of a relational approach to biology has, in the last two
decades, become quite influential under the name of ``network
science''---we refer to the recent textbook by \citet{Bar:16a}, one of
the most influential figures of the discipline, as an introduction to
the large body of literature. Papers in network science often follow
statistical approaches for inferring networks from data and the
resulting networks are analysed by computational techniques developed
with methods from graph theory and statistical mechanics.

Studies in the area of network science clearly illustrate the
challenges with ``throwing away the physics and keeping the
organisation''. For example, once a large-scale biological network has
been inferred from data, the question of its interpretation might not
be easy to answer, precisely because only minimal assumptions are made
regarding the properties of individual nodes. Is it most important
that a network has certain global properties such as being scale-free
\citep{Wat:98a, Bar:99a}, to which extent the network can be
controlled \citep{Liu:11a, Rut:14a} or that certain ``network motifs''
are more prevalent than expected by chance \citep{Mil:02a, Alo:07a}?
If we consider, more specifically, gene regulatory networks, another
problem becomes apparent. Rosen always assumes that relationships
between the components of his~\mrsys\ are known. Unfortunately,
inferring the interactions within biochemical networks such as the
highly complicated large networks of genes and their transcripts is
often quite challenging and might not lead to conclusive results. This
motivated, for example, \citet{Oat:14a} to include a more detailed
model of the underlying reactions in order to obtain more accurate
results on the interactions between the components of the biochemical
network. Also, as far as the impact of the conclusions is concerned,
one might argue that combining network models with at least some
description of the underlying ``physics'' of the components has been
more promising than studies that are restricted to networks whose
nodes without considering their underlying structure. As an example I
refer to \citet{Col:06a} who investigated the global spread of
epidemics along the airline traffic network using the example of the
2002 outbreak of severe acute respiratory syndrome (SARS).


In summary, by considering the example of ``network science'' which is
the area of science that is probably most closely related to Rosen's
idea of a relational biology, the benefit of Rosen's proposal of
throwing away the physics and keeping the organisation are not
entirely clear. But this has to be considered in a situation where
applications of his ideas are still in relatively early stages because
Rosen himself did not work towards applying his theory to concrete
biological problems. 

{
\subsection{Rosen's relational biology and category theory}
\label{sec:categorytheory}

As mentioned several times above, a strong view of Rosen's is the
motto ``throw away the physics and keep the organisation''. For this
reason he deliberately defines the components of his \mrsys\ as
``black boxes''. Rather than describing the structure of components
that form a system, his relational biology focuses on their
interactions with other components. Consistently, he primarily uses
category theory for describing interactions between objects; that the
objects are members of categories with certain underlying mathematical
structures hardly plays any role.

This use of category theory is likely to disappoint most readers of
Rosen's works with a mathematical background. In the preface of her
recent textbook ``Category Theory in Context'' Emily Riehl introduces
category theory like this \citep{Rie:16a}:

\begin{quotation}
  Atiyah\footnote{Sir Michael Atiyah (*1929), British mathematician,
    one of the most distinguished mathematicians of the 20th
    century. Apart from many other awards and honours he won the
    Fields medal (1966) and the Abel Prize (2004).} described
  mathematics as the ``science of analogy.'' In this vein, the purview
  of category theory is \emph{mathematical analogy}. Category theory
  provides a cross-disciplinary language for mathematics designed to
  delineate general phenomena, which enables the transfer of ideas
  from one area of study to another.
\end{quotation}

A strong motivation for the development of category theory and one of
the main reasons for its success as a mathematical discipline is the
formalisation of links between different areas of mathematics. When
category theory was originally formulated by \citet{Eil:45a} the new
notions of category theory facilitated understanding the connections
between topological spaces and algebraic objects such as groups or
vector spaces that can be associated to them. The ability for finding
such mathematical analogies (between topological and algebraic objects
in the case of algebraic topology) crucially depends both on relations
between objects from particular categories (such as topological spaces
or groups) but also on the underlying mathematical structures of these
objects that are preserved by morphisms.

\subsubsection{Rosen's treatment of category theory as an incremental extension of graph theory}
\label{sec:criticalCT}

But, as already mentioned, Rosen explicitly avoids assigning structure
to the components of his \mrsys. From the introduction of
\citep{Rosen:58b} it becomes clear that Rosen regards category theory
as an incremental extension of graph theory that enables him to more
flexibly describe relations between ``black boxes''.  Unfortunately,
this prevents him from taking much advantage from the main strength of
category theory, namely, relating the structure of mathematical
objects appearing in different disciplines of
mathematics. 

The negative impact of this use of category theory on an audience of
applied mathematicians must not be underestimated. A strong motivation
in mathematics itself as well as in the community of applied
mathematicians is to use mathematical notions as efficiently as
possible. By failing to take full advantage of the ability of category
theory to relate mathematical structures Rosen does not only miss the
chance to capture properties of a biological system that might be
encoded in such structures. Even worse, an audience from a
mathematical background might even be deterred from Rosen's ideas, not
because of the ideas themselves but due to the perceived shortcomings
in their mathematical presentation. In summary, one might go as far as
saying that instead of being a strength of Rosen's theories, category
theory is one of the most important obstacles for their
acceptance. Nevertheless one shall not be overly critical of Rosen's
approach of using category theory for his research---after all he was
a pioneer in applying a novel, quite difficult mathematical discipline
at a time when even the foundations of this discipline were still
under development.

\subsubsection{An alternative perspective on relational biology}
\label{sec:alternativeCT}

Does the preceding section imply that category theory is not an
appropriate tool for investigating biological systems? One reviewer of
an earlier version of this manuscript kindly directed me to work from
the group of mathematical physicist John Baez who have most recently
applied category theory to open reaction networks \citep{Bae:17a}. It
is worthwhile to compare this emerging research with \mrsys\ which
were inspired by networks of metabolic reactions. Rather than
abstracting reaction networks to a network of input-output
relationships, Baez and co-workers go the opposite way---they develop
specific categories \textbf{RNet} and \textbf{RxNet} in order to
formalise Petri Nets, a diagrammatic representation of chemical
reactions. By constructing a functor to the category \textbf{Dynam} of
open dynamical systems they can relate a given Petri net to a system
of ordinary differential equations, the rate equations associated with
this particular Petri net. A functor from \textbf{Dynam} to the
category of relations \textbf{Rel} maps dynamical systems to their
steady states. Of course, the ambition of Baez and co-workers is
presumably not to answer questions like ``What is life?''  but their
``compositional framework'' allows them to build reaction networks
from simpler components via composition of morphisms and relate the
structural properties of reaction networks to similar models such as
electrical circuits \citep{Bae:16a}, signal-flow diagrams
\citep{Erb:16a} and Markov processes \citep{Bae:16b} which are all
formalised in a similar way as described above using the language of
category theory. It seems clear that although this work is not less
``relational'' than Rosen's \mrsys\ or the network science approaches
mentioned in Section~\ref{sec:discussWhat}, the publications from
Baez's group clearly take advantage of category theory for relating
the different mathematical structures characteristic of different
modelling approaches.


\subsection{Rosen's modelling relation and mechanistic models in
  applied mathematics}
\label{sec:discussRosenAppliedM}

In ``Life Itself'', \citet{Rosen:91a} repeatedly proposes relational
models as alternatives to mechanistic models. But, first of all,
according to the view explained in Section~\ref{sec:model},
mechanistic models should not be considered as the opposite of
relational models, in fact, relational models can be regarded as a
particular type of mechanistic model. Second, many of Rosen's
objections arise because he implicitly assumes that mechanistic models
necessarily have to be interpreted in a reductionistic way. But
building a mechanistic model does not mean that the natural system
necessarily must be ``reduced'' to the mechanism represented by the
model---in fact, this is just a specific interpretation. In contrast,
in Section~\ref{sec:model} we propose an alternative perspective on
mechanistic models---according to this view the ``mechanism''
represented in a mechanistic model only provides an explanation of a
particular aspect of the system behaviour which is defined by the
underlying assumptions of the model. If additional aspects of the
system behaviour are to be considered this requires refining the
assumptions of the model, as a result the new model will represent a
more detailed mechanism that provides a more comprehensive (but still
partial) explanation of the system behaviour.

\subsection{Rosennean complexity and mechanistic models}
\label{sec:rosencomplexity}

The difference between Rosen's view on modelling and the view I
outlined in Section~\ref{sec:model} is most obvious when considering
his definition of complex systems:

\begin{quotation}
  A system is simple if all its models are simulable. A system that is
  not simple, and that accordingly must have at least one
  non-simulable model, is complex.
\end{quotation}

Rosen's concept of complexity is a direct consequence of his modelling
relationship. With his modelling relationship he outlines an approach
that enables us to directly relate a natural system to a formal
system, the model. Thus, it might seem that modelling is, at least
conceptually, trivial. Indeed, for our example of an algebraic
curve~$\mathcal{C}$ (Section~\ref{sec:algebraiccurves}), the
representation of the functor~$\mathcal{C}$ by the coordinate
ring~$A$~\eqref{eq:coordinatering} provides us with a ``model'' for
the algebraic curve over arbitrary~$K$-algebras. But according to
Rosen's definition, the fact that the coordinate ring~$A$ exists as a
model for arbitrary~$K$-algebras makes algebraic curves a ``simple''
system. For a ``complex'' system, an analogue of the coordinate
ring~$A$ might still exist but it is ``non-simulable'' which Rosen
defines as not Turing-computable. Although such a model would still
perfectly describe the natural system, the required calculations
formalised by a Turing machine might not terminate in finite time.

Most striking about Rosen's definition of a complex system is how the
relationship of a system and ``its'' models is described. According to
Section~\ref{sec:model} and in contrast to Rosen's concept of the
modelling relation a system does not ``have'' models---models cannot
be objectively associated with a system via the modelling relationship
but rather are subjectively attributed to the system by the
modeller. A model will---due to the requirement of making simplifying
assumptions---necessarily always remain incomplete, the case of a
``simple'' system in Rosennean terms, where a perfect formal
representation of a natural system can be found, does not
exist. Mechanistic models serve a specific purpose by efficiently
representing a set of underlying assumptions that are consistent with
the purpose of the model. As explained in Section~\ref{sec:model},
rather than being a formal representation of a particular ``truth''
about a natural system, the aim of mechanistic models is to explore
scientific hypotheses that are represented in the assumptions of the
model. Therefore, consideration of competing models based on
alternative assumptions is an important part of scientific discussion
in the literature.

\subsection{Rosennean complexity in the toolbox of mathematical
  biologists}
\label{sec:discussrelationalbiology}

The most important aspect of Rosen's theories is his postulate that
``organisms are closed to efficient causation''. In order to to assess
how well this notion is able to describe living systems one has to go
beyond theoretical considerations. From the perspective of a modeller
it is therefore the most important shortcoming that so far there are
very few examples of mathematical models implementing Rosen's
postulate in the context of concrete biological systems.

In the field of modelling biochemical reactions, this issue is being
addressed by the group of C\'{a}rdenas and Cornish-Bowden. In a series
of papers, \citet{Let:06a,Cor:07a, Cor:07b} developed a simple \mr\
system representing a simple biochemical reaction network that was
later implemented in a mathematical model and investigated by
simulation \citep{Pie:10a, Pie:12a, Pie:12b,Cor:13b}. 

Whereas organisms are postulated to be closed to efficient causation,
they must be open to ``material causes'' as explained in
Section~\ref{sec:complex} or, in more familiar terms, flows of matter
and energy. A mathematical model that realises Rosen's concept of an
organism should therefore also appropriately account for the exchange
of matter and energy with the environment. The bond graph methodology
\citep{Bor:10a} is an established approach from the engineering
literature for building models of complex systems with energy flows
between multiple domains (electrical, chemical etc.). Bond graphs were
recently applied to biochemical reactions \citep{Gaw:14a,Gaw:15a} and
subsequently to more complex physiological systems
\citep{Gaw:16b,Gaw:17a}. Also, in mathematical ecology there are
several examples of energy-based models ranging from the early
$(E, M)$ framework developed by \citet{Sme:76a} to the more recent
studies by \cite{Cro:12a} and \citet{Bat:15a}. Extending these
frameworks by modelling approaches that realise closure to efficient
causation will enable us to investigate the significance of Rosennean
Complexity for ecosystems. 

{In all implementations of Rosen's principle of closure to efficient
  causation an obvious difficulty is to identify ``efficient causes''
  and distinguish them from ``material causes''---Rosen's own
  publications provide relatively little guidance due to the mostly
  formal presentation with very few specific biological examples. In
  this regard, the recent paper \citet{Mos:16a} is highly relevant:
  the authors develop a theory of biological organisation that
  comprises Rosen's views but draws from the much longer tradition of
  organicism. Organicism is a perspective on biology which states that
  ``organisms are the main object of biological science because [...]
  they cannot be reduced to more fundamental biological entities (such
  as the genes or other inert components of the organism)''
  \citep{Mos:16a}. Organicism implies that the individual parts that
  the organism consists of can only be understood by taking into
  account their relationships and interactions with other parts. But
  crucially, in contrast to the approach that Rosen takes with his
  \mrsys, \citet{Mos:16a} avoid reducing the parts of an organism to
  ``black boxes'' without underlying structure. Instead, they identify
  specific parts of biological systems as \emph{constraints} which
  control processes without themselves being altered by them---they
  give the role that enzymes play in reaction networks and the
  influence of the vascular system on the flow of oxygen in the body
  as examples. Biological organisation according to \citet{Mos:16a} is
  realised via ``closure of constraints'' which shows that constraints
  are a closely related concept to Rosen's ``efficient causes''. This
  idea of biological organisation is one of three theoretical
  principles for biology proposed in the highly readable special issue
  ``From the century of the genome to the century of the organism: New
  theoretical approaches'' published in \emph{Progress in Biophysics
    and Molecular Biology}---the others are variation \citep{Mon:16a}
  and a postulated biological ``default state'' \citep{Sot:16b}. The
  articles from this special issue provide valuable guidance to
  modellers who wish to construct models which represent Rosen's idea
  of an organism or, indeed, stand in the much longer tradition of
  organicism.}

Although it seems clear that the task of building such mathematical
models that provide a better representation of organisms will not be
an easy one, it seems equally clear that this will bring us another
step further towards answering the grand question ``What is Life?''.

\section*{Acknowledgements}
I cordially thank two anonymous reviewers for their thoughtful and
supportive reviews which greatly improved this article. I am
particularly grateful to both reviewers for directing me to highly
relevant additional literature.

 \bibliographystyle{/Users/merlin/latex/style/elsart-harv}
\bibliography{/Users/merlin/Documents/references/refer}

\end{document}